\documentclass[a4paper]{VisionStyle}
\usepackage{epsfig}

\begin{document}

\title{Recent CHANDRA and XMM observations of two very different galaxy clusters: \object{RBS797} and \object{CL 0939+4713}}

\author{E.\,De Filippis\inst{1} \and S.\,Schindler\inst{1} \and A.\,Castillo-Morales\inst{1,2}} 

\institute{
Astrophysics Research Institute, Twelve Quays House, Egerton Wharf, CH41 1LD Birkenhead, Wirral UK
\and
Universidad de Granada, Dpto. Fisica Teorica y del Cosmos, Avda. Fuentenueva, s/n 18002 Granada, Spain
}

\maketitle 

\begin{abstract}
\noindent \object{RBS797} and \object{CL 0939+4713} are two intermediate red-shift clusters ($z=0.35-0.41$). They have very different morphologies but both show surprisingly interesting structures. \object{RBS797} looks relaxed, with an almost circular morphology; a CHANDRA observation of this cluster has revealed two deep depressions in the X-ray emission near the core. \object{CL 0939+4713 } has instead an irregular morphology with evident substructures which seem to be in the process of merging.\\
Throughout this talk, if not otherwise stated, the errors are $90\%$ confidence level.\\

\end{abstract}

\section{Introduction}
The latest results from the analysis of a CHANDRA observation of the hot ($T=7.7$~keV), distant ($z=0.35$) galaxy cluster \object{RBS797} will first be presented (see \S~\ref{edefilippis-B3_sec:RBS797}). The most striking observed features are two deep depressions in the X-ray emission in the core of the cluster. This is the first time such depressions have been observed in a "distant" cluster. The low temperature of the high density regions surrounding the holes is a clear indication that these cannot be shock regions. It is likely that the intra-cluster gas has been subsonically pushed away from the areas of low X-ray emission to the areas of higher emission by the pressure of the relativistic particles in radio lobes.\\
\noindent The first results from a recent XMM observation of the galaxy cluster (\object{CL 0939+4713} - \object{A851}) will then be presented (see \S~\ref{edefilippis-B3_sec:Cl0939}). \object{CL 0939+4713} lies at a slightly larger distance than \object{RBS797} ($z=0.41$) but its complex and irregular structure is completely different from the almost spherical and relaxed one of \object{RBS797}.\\
Its high richness, together with the presence of numerous small, blue objects in the proximity of a background quasar ($z=2.05$), has made \object{CL 0939+4713}, in the past decade, a fundamental target in the study of the evolution of galaxy clusters.\\
Previous ROSAT observations have shown the cluster complex internal structure. The ROSAT analysis left, though, high uncertainties in the determination of the physical conditions of the intra-cluster gas and in the identification of its substructures.\\

\section{\object{RBS797}}
\label{edefilippis-B3_sec:RBS797}
\object{RBS797} was discovered during the ROSAT Bright Survey (\cite{edefilippis-B3:Schw00}). It is a distant (with a spectroscopic red-shift of $z=0.35$) and luminous cluster (with an X-ray luminosity of $L_{X}=7 \times 10^{45}$~erg/s).\\
We have observed \object{RBS797} with CHANDRA (ACIS-I for a total exposure time of $13.3$~ksec) as part of a large programme to search for gravitational arcs in clusters.\\
The goal of the project is to take both deep optical and high resolution X-ray observations, in order to use these clusters for arc statistics and to compute and compare the cluster masses estimated with X-rays and with gravitation lensing. To this aim, our cluster sample has been extracted in the X-rays from the ROSAT Bright Survey, in order to select the most luminous and massive galaxy clusters observed in the ROSAT All Sky Survey.\\ 
Unfortunately the optical data are not available yet. In this talk we will present the results of the X-ray observation.\\

\subsection{General morphology}
\begin{figure}
  \begin{center}
    \epsfig{file=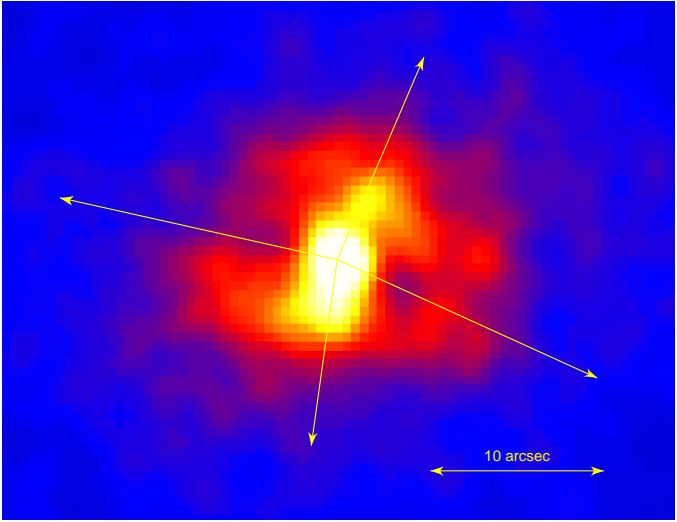, width=9cm}
  \end{center}
\caption{Smoothed image of the core of the cluster \object{RBS797}. The arrows point the directions along which the four surface brightness profiles were taken (see Fig.~\ref{edefilippis-B3_fig:profiles}).}  
\label{edefilippis-B3_fig:ima797}
\end{figure}

\begin{figure}
  \begin{center}
    \epsfig{file=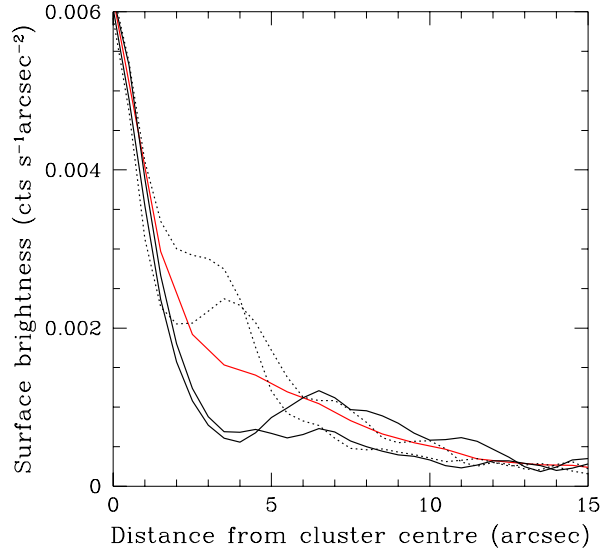, width=9cm}
  \end{center}
\caption{Surface brightness profiles in direction of the minima (solid lines), and in direction of the maxima (dotted lines). The red line shows an average profile integrated over all angles.}
\label{edefilippis-B3_fig:profiles}
\end{figure}

At large radii \object{RBS797} shows a very regular morphology; it looks fairly relaxed, with a slight elliptical shape. Fig.~\ref{edefilippis-B3_fig:ima797} shows a smoothed image of the core of the cluster, which has instead a peculiar structure, with two strong depressions in the X-ray emission. These depressions are almost circular, with a radius of $\approx 30$~kpc, and lie in opposite directions respect to the cluster centre, at a distance of about $40$~kpc from it.\\
\noindent In order to quantify the depth of these depressions we took four surface brightness profiles, two in direction of the voids and two in the perpendicular directions; the result is plotted in Fig.~\ref{edefilippis-B3_fig:profiles}. The two profiles in direction of the voids (solid lines) show strong and steep minima, while the trend is completely different in the perpendicular directions (dotted lines) and for the average profile computed over the whole cluster (red solid line): in correspondence of the voids, the surface brightness becomes up to $3-4$ times fainter than in the other directions.\\
\noindent These kind of depressions in the X-ray emission have recently been observed also in a few other clusters (\cite{edefilippis-B3:Fab99}, \cite{edefilippis-B3:McN00}, etc.). \object{RBS797} is, though, a special case since it has given the chance to observe for the first time these features in a relatively distant cluster and, furthermore, in this cluster the depressions are much deeper than the ones observed in any other cluster until now.\\
The origin of these depressions is today still not sure. One of the most realistic hypothesis is that these features are due to the presence of radio lobes; inside these lobes the relativistic particles would push away the outside intra-cluster gas, slowly compressing it on the border of the lobes that subsequently become high density regions.\\

\subsection{Radio observations}
To verify the presence, in the core of the cluster, of a diffuse radio emission in the positions where the X-ray emission seems to be suppressed, together with Luigina Feretti we have applied and obtained time at the VLA in D configuration.\\
\begin{figure*}[ht]
  \begin{center}
    \epsfig{file=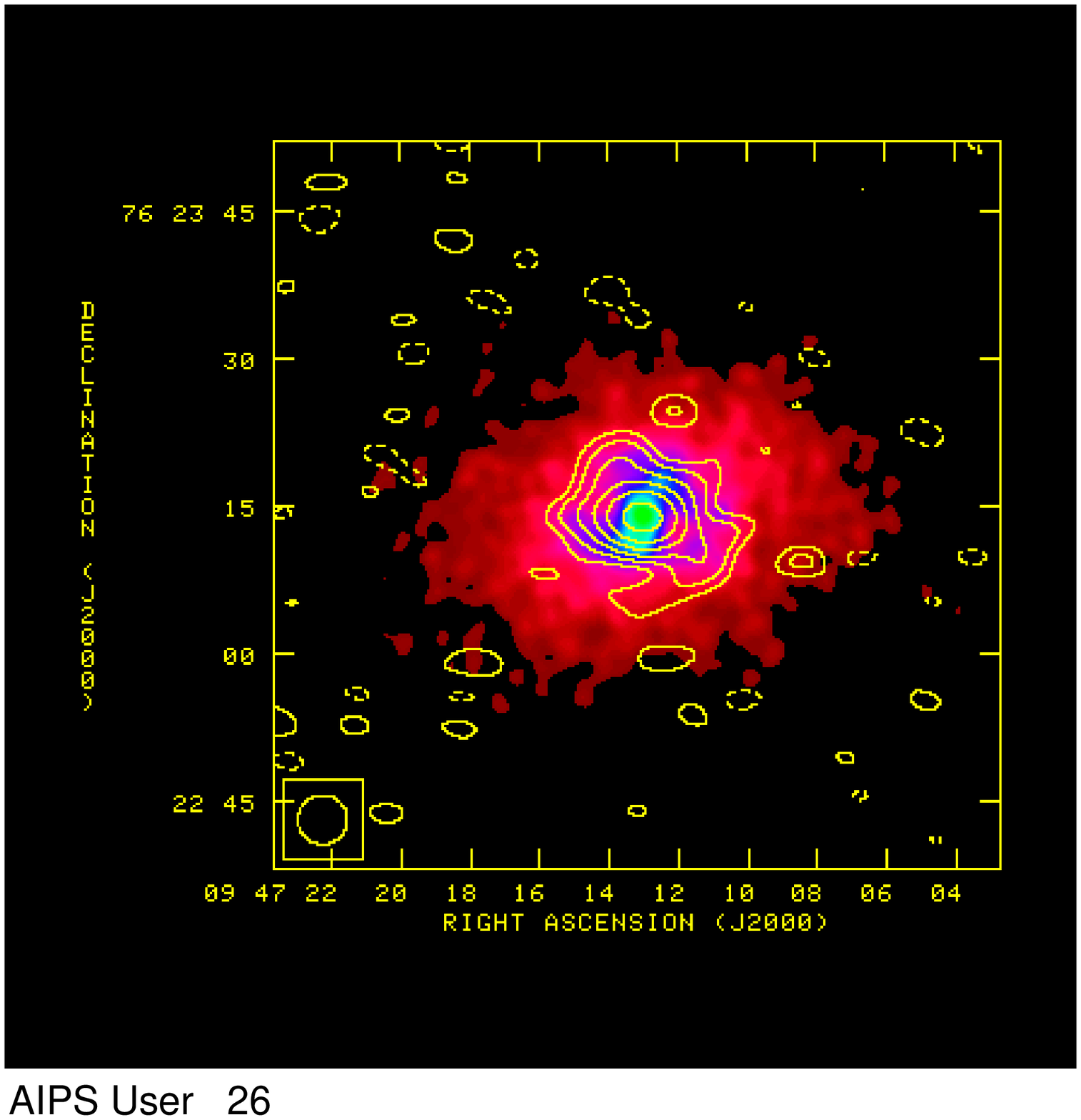, width=18cm}
  \end{center}
\caption{Radio contours superimposed on the smoothed X-ray colour image of the core of \object{RBS797}.}  
\label{edefilippis-B3_fig:radio}
\end{figure*}
A first result of the analysis of this radio observation is shown in Fig.~\ref{edefilippis-B3_fig:radio} where the radio contours are superimposed on the coloured smoothed X-ray image of the core of the cluster.\\
\noindent The radio data confirm the presence of a strong radio source positioned in the centre of the cluster. The detected emission differs from the one coming from a regular point source, showing an extension in the same direction of the depressions observed in the X-rays; nothing definite yet can unfortunately be said because of the low resolution of the data (due to the configuration of the interferometer for this observation).\\
To obtain radio data with the necessary resolution we have now applied for more VLA time, but this time with the interferometer in the A+B configurations. This new observation will finally allow us to obtain the required resolution to definitively verify the presence of a diffuse radio emission and, in such a case, to confirm if this emission is in coincidence of the X-ray depressions, giving the chance to put strong constrains on the interaction between the intra-cluster gas and the relativistic particles inside the radio lobes.\\

\subsection{Spectral analysis and mass determination}
\begin{figure}
  \begin{center}
    \epsfig{file=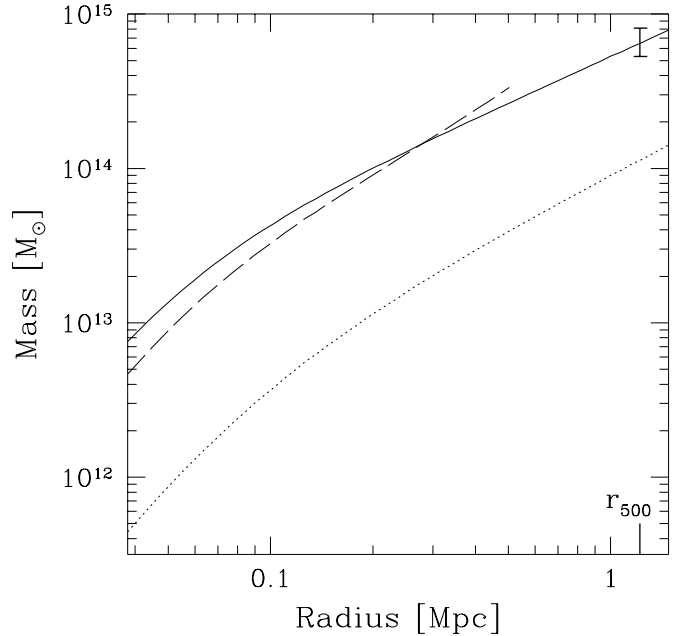, width=9cm}
  \end{center}
\caption{Integrated mass versus radius: total gravitational mass (solid line) and gas mass (dotted line). A typical error bar is shown at $r_{500}$. The total mass profile, taking into account a temperature gradient (dashed line), is shown up to the radius to which the temperature gradient can be determined.}
\label{edefilippis-B3_fig:mass}
\end{figure}

An analysis of the spectral properties of the intra-cluster gas has been performed. The emission coming from the central source and other point sources present in the field of view of the cluster have been excluded. The resulting diffuse emission, coming from the ICM, is well fitted by a MEKAL model, which gives an average temperature for the whole cluster of $T=7.7_{-1.0}^{+1.2}$~keV and a metal abundance of $Z=0.26\pm0.1\ Z_{\rm \odot}$.\\
A radial temperature profile for the cluster has also been computed. For this purpose, the emission coming from the cluster has been split in three concentric annular regions. The emission detected inside each annulus has been independently fitted with a MEKAL model, giving the following values for the temperature and the metal abundance:
\begin{eqnarray}
2\arcsec-10\arcsec: \, \, \, T=5.7_{-0.5}^{+0.7}\ {\rm keV}; Z=0.38_{-0.16}^{+0.17}\ Z_{\rm \odot}
\label{edefilippis-B3_eq:eq1}\\
10\arcsec-30\arcsec: \, \, \, T=8.4_{-1.0}^{+1.2}\ {\rm keV}; Z=0.25_{-0.18}^{+0.18}\ Z_{\rm \odot}
\label{edefilippis-B3_eq:eq2}\\
30\arcsec-80\arcsec: \, \, \, T=11.7_{-2.5}^{+4.3}\ {\rm keV}; Z=0.20_{-0.20}^{+0.35}\ Z_{\rm \odot}
\label{edefilippis-B3_eq:eq3}
\end{eqnarray}
\begin{figure}
  \begin{center}
    \epsfig{file=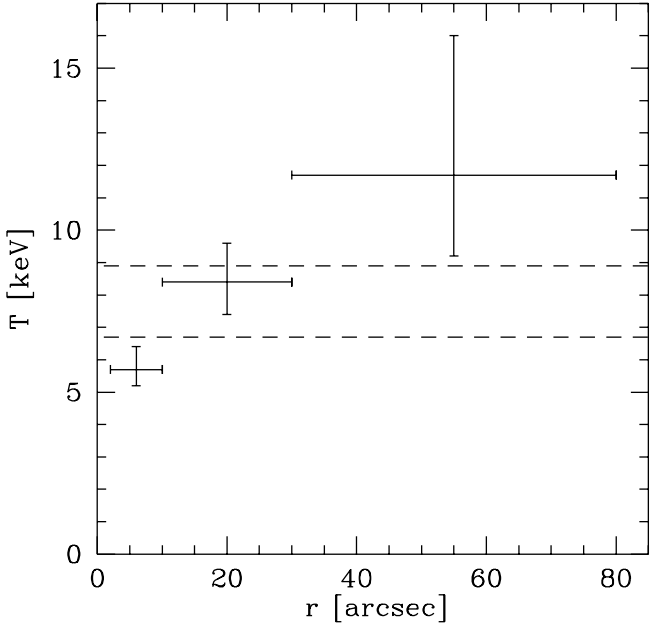, width=9cm}
  \end{center}
\caption{Temperature profile of \object{RBS797}. The dotted lines give the $90\%$ confidence level of the average ambient temperature of the cluster.}
\label{edefilippis-B3_fig:Tprofile}
\end{figure}
Eqs.~\ref{edefilippis-B3_eq:eq1}, \ref{edefilippis-B3_eq:eq2} and \ref{edefilippis-B3_eq:eq3} show a slight trend of increasing temperature with the radius, and decreasing values of the metallicity; this trend can also be observed in Fig.~\ref{edefilippis-B3_fig:Tprofile}.\\
We have also attempted a spectral fit of the X-ray emission coming from the ``arms'', the regions of higher emission surrounding the minima. In these regions we find a temperature ($T=4.4_{-0.6}^{+0.7}$~keV) that is slightly lower than for the rest of the cluster. This result suggests that the ``arms'' are not shocked regions, but that the radio lobes are probably rising by buoyancy (\cite{edefilippis-B3:McN00}) in the ICM compressing, on their borders, gas that was previously in the cluster centre.\\
We have also performed a mass determination for this cluster (see Fig.~\ref{edefilippis-B3_fig:mass}). At a radius of $r_{500}$ (where $r_{500}=1.22$~Mpc) we find, with the assumption of spherical symmetry, a value for the gas mass of $M_{\rm gas}(r_{500})=1.13\pm0.09\times 10^{14}M_{\rm \odot}$. With the further assumption that the gas in the cluster is in hydrostatic equilibrium, we find that the total mass of the cluster is $M_{\rm tot}(r_{500})=6.5^{+1.5}_{-1.2}\times 10^{14}M_{\rm \odot}$. This corresponds to a gas mass fraction: $f_{\rm gas}(r_{500})=0.17\pm0.05$.\\

\section{\object{CL 0939+4713}}
\label{edefilippis-B3_sec:Cl0939}
Cluster \object{CL 0939+4713} is at a red-shift of $z=0.407$ and is a very luminous galaxy cluster in the optical waveband. Contrarily to \object{RBS797}, \object{CL 0939+4713} is a very ``popular'' cluster; it is a famous Butcher-Oemler cluster and it has been observed at many wavelengths and by many different instruments.\\
We have now observed it with XMM (EPIC MOS 1 \& 2 and PN) for a total exposure time of $47$~ksec\footnote{M.P.E. guaranteed time: TS Bernd Aschenbach}.\\
\begin{figure*}
 \begin{center}
   \epsfig{file=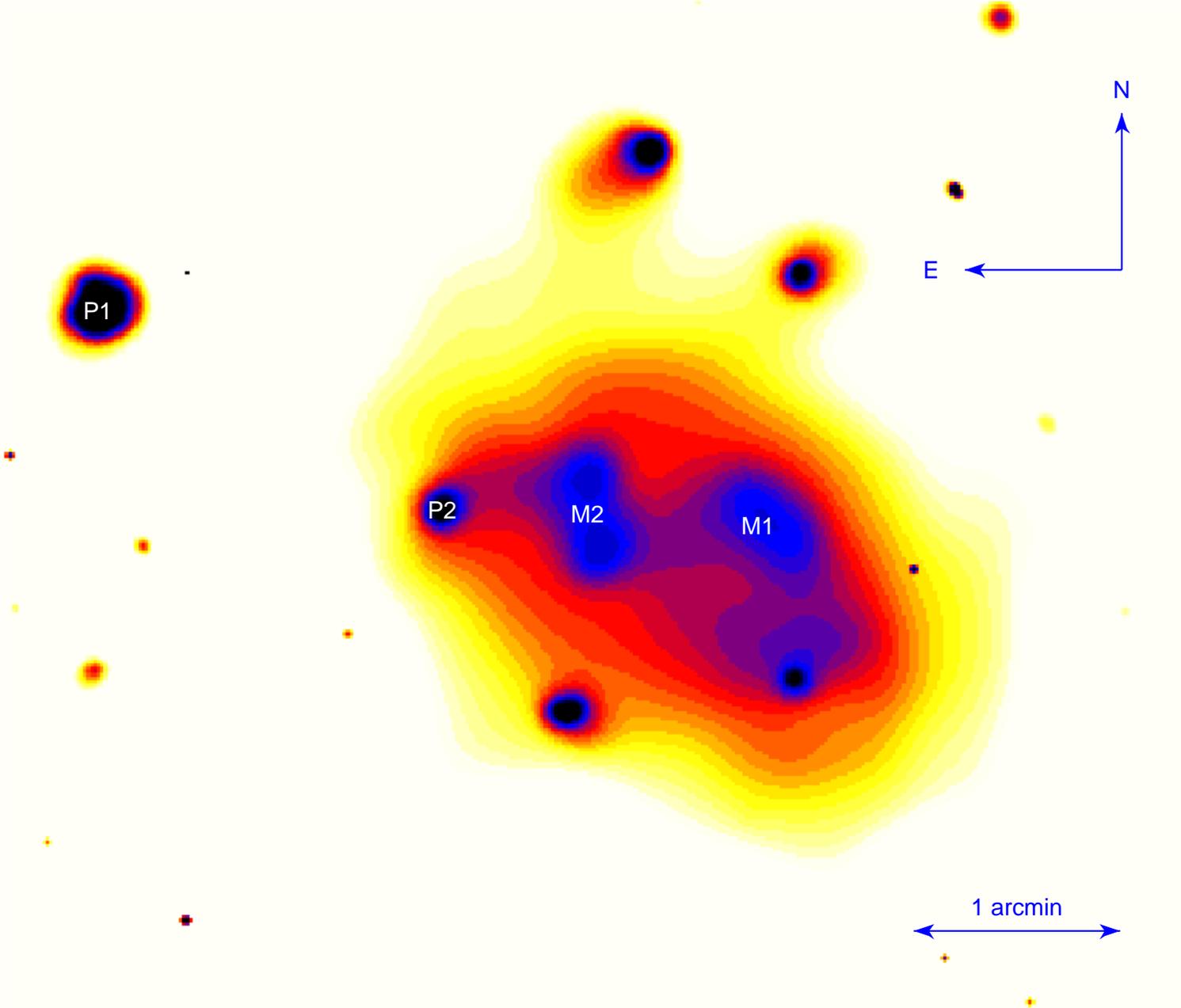, height=20cm, angle=270}
 \end{center}
\caption{Smoothed image of the core of \object{CL 0939+4713}: MOS1, MOS2 \& PN superimposed.}
\label{edefilippis-B3_fig:ima0939}
\end{figure*}
Fig.~\ref{edefilippis-B3_fig:ima0939} shows a smoothed X-ray image of the centre of the cluster, where the three sets of data taken with the MOS1, MOS2 and PN have been superimposed. The most evident features are two main substructures (M1 \& M2) that compose the centre of the cluster. P1 is a foreground source; P2 is a quasar at $z=2$.\\

\subsection{Preliminary spectral analysis}
Thanks to this XMM observation, we have been able for the first time to resolve many of the point sources which lie in the field of view of the cluster. This allows to isolate their emission and to obtain a much better analysis of the physical properties of the cluster ICM, if compared to previous X-ray observations.\\
We have just started to perform a spectral analysis for this cluster. After having subtracted the background and the emission coming from all the point sources, and having corrected for vignetting, the diffuse emission coming from the cluster is well fitted by a MEKAL model. The spectral fit gives an average value for the cluster temperature of $T_{(0.5-10.0\ {\rm keV})}=4.9\pm 0.3$~keV and a value for the metal abundance of $Z=0.18^{+0.08}_{-0.07}Z_{\rm \odot}$.\\
The cluster shows a low luminosity in the X-rays ($L_{\rm X}({\rm bol})=8.9 \times 10^{44}$~erg/s), which was not expected from such an optically rich and luminous cluster. The reason could probably be due to the presence of the two main substructures that still have to merge.\\
In order to try to obtain a better understanding of the physical properties of this cluster, we have also independently analysed the spectral properties of the diffuse emission from the two subregions {\it M1} and {\it M2} and from the central region {\it C}, lying between {\it M1} and {\it M2}. From a preliminary analysis we observe a trend of increasing metal abundance going from region {\it M2} to region {\it M1} (see Eqs.~\ref{edefilippis-B3_eq:eq4}, \ref{edefilippis-B3_eq:eq5} and \ref{edefilippis-B3_eq:eq6}).
\begin{eqnarray}
 \, \, \, \, \, \, \, \, \, M2: \, \, \, \, \, \, \, \, \, Z=0.18_{-0.18}^{+0.15}\ Z_{\rm \odot}
\label{edefilippis-B3_eq:eq4}\\
 \, \, \, \, \, \, \, \, \, C:  \, \, \, \, \, \, \, \, \, Z=0.31_{-0.19}^{+0.20}\ Z_{\rm \odot}
\label{edefilippis-B3_eq:eq5}\\
 \, \, \, \, \, \, \, \, \, M1: \, \, \, \, \, \, \, \, \, Z=0.32_{-0.18}^{+0.16}\ Z_{\rm \odot}
\label{edefilippis-B3_eq:eq6}
\end{eqnarray}
An extensive red-shift survey has been performed in the past years on the galaxies lying the field of view of this cluster (\cite{edefilippis-B3:Dre99}). More than $60$ of them have been proved to lie at approximately the same distance from us; a large number of the biggest and most luminous of these galaxies lies inside the subregion {\it M1}. These galaxies could be strongly enriching the intra-cluster gas with heavy metals, and this could explain the observed trend of increasing metal abundances.\\
We also find that the two subregions {\it M1} and {\it M2} have the same temperature, which is slightly lower than the one found in the central region {\it C} (see Eqs.~\ref{edefilippis-B3_eq:eq7}, \ref{edefilippis-B3_eq:eq8}, \ref{edefilippis-B3_eq:eq9}). This could be a sign that the two subregions are in the process of merging, and therefore beginning to compress and heat the ICM between them.
\begin{eqnarray}
 \, \, \, \, \, \, \, \, \, M2:  \, \, \, \, \, \, \, \, \, T=5.1_{-0.5}^{+0.8}\ {\rm keV}
\label{edefilippis-B3_eq:eq7}\\
 \, \, \, \, \, \, \, \, \, C:   \, \, \, \, \, \, \, \, \, T=5.4_{-0.5}^{+0.6}\ {\rm keV}
\label{edefilippis-B3_eq:eq8}\\
 \, \, \, \, \, \, \, \, \, M1:  \, \, \, \, \, \, \, \, \, T=5.1_{-0.4}^{+0.5}\ {\rm keV}
\label{edefilippis-B3_eq:eq9}
\end{eqnarray}
Unfortunately, for both the metal abundance and temperature profiles, the values for the different regions are almost consistent within their errors. We state again that these are just preliminary results and that in the near future, thanks to a new and more accurate calibration of the data, we should be able to get much better results.\\

\section{Conclusions and further work}
We have observed two intermediate red-shift clusters with the two new X-ray satellites: CHANDRA and XMM.\\
\object{RBS797} looks almost relaxed,  \object{CL 0939+4713} has two substructures which seem to be just in process of merging.\\

\noindent With CHANDRA, thanks to its amazing angular resolution, in a recently discovered cluster such as \object{RBS797}, we have been able to detect X-ray depressions, even in such a distant cluster; this has allowed us a straightforward prediction of the presence, in the centre of the cluster, of radio lobes generated from the central active galaxy. This prediction has been partially confirmed by low resolution radio data. New radio data, which we have now applied for, will then help us in a better understanding of the interaction between the relativistic particles and the particles in the intra-cluster gas.\\

\noindent With XMM we have instead analysed a cluster which had been previously extensively studied and observed. We have resolved two main substructures forming the cluster centre and many point sources within the cluster. This is giving us the chance to perform a very accurate spectral analysis for this cluster that will allow us a study of the physical conditions of the intra-cluster gas.\\

\end{document}